\documentclass{emulateapj}
\usepackage{times}
\usepackage{graphicx, color}% Include figure files
\usepackage{hyperref}
\usepackage{lineno}

\slugcomment{Not to appear in Nonlearned J., 45.}

\shorttitle{HESS J1640$-$465 and HESS J1641$-$463 with \emph{Fermi}-LAT}
\shortauthors{Lemoine-Goumard et al. 2014}

\begin{document}

\title{HESS J1640$-$465 and HESS J1641$-$463: two intriguing TeV sources in the light of new \emph{Fermi} LAT observations}


\author{
M.~Lemoine-Goumard\altaffilmark{1,2}, 
M.-H.~Grondin\altaffilmark{1,3}, 
F.~Acero\altaffilmark{4}, 
J.~Ballet\altaffilmark{4}, 
H.~Laffon\altaffilmark{1}, 
T.~Reposeur\altaffilmark{1}
}
\altaffiltext{1}{Centre d'\'Etudes Nucl\'eaires de Bordeaux Gradignan, IN2P3/CNRS, Universit\'e Bordeaux 1, BP120, F-33175 Gradignan Cedex, France}
\altaffiltext{2}{email: lemoine@cenbg.in2p3.fr}
\altaffiltext{3}{email: grondin@cenbg.in2p3.fr}
\altaffiltext{4}{Laboratoire AIM, CEA-IRFU/CNRS/Universit\'e Paris Diderot, Service d'Astrophysique, CEA Saclay, 91191 Gif sur Yvette, France}

\begin{abstract}
We report on $\gamma$-ray analysis of the region containing the bright TeV source HESS J1640$-$465 and the close-by TeV source HESS J1641$-$463 using 64 months of observations with the \emph{Fermi} Large Area Telescope (LAT). Previously only one GeV source was reported in this region and was associated with HESS J1640$-$465. With an increased dataset and the improved sensitivity afforded by the reprocessed data (P7REP) of the LAT, we now report the detection, morphological study and spectral analysis of two distinct sources above 100 MeV. The softest emission in this region comes from the TeV source HESS J1641$-$463 which is well fitted with a power law of index $\Gamma = 2.47 \pm 0.05 \pm 0.06$ and presents no significant $\gamma$-ray signal above 10 GeV, which contrasts with its hard spectrum at TeV energies. The \emph{Fermi}-LAT spectrum of the second TeV source, HESS J1640$-$465 is well described by a power-law shape of index $\Gamma = 1.99 \pm 0.04 \pm 0.07$ that links up naturally with the spectral data points obtained by the High Energy Stereoscopic System (H.E.S.S.). These new results provide new constraints concerning the identification of these two puzzling $\gamma$-ray sources.
\end{abstract}

\sloppy

\section{Introduction}

HESS J1640$-$465 is an extended (2.7 $\pm$ 0.5 arcmin) $\gamma$-ray source detected by the High Energy Stereoscopic System (H.E.S.S.) during its survey of the Galactic Plane \citep{galplane}. It is centered within the radio supernova remnant (SNR) G338.3$-$0.0 \citep{whiteoak96}. Using X-ray observations with \emph{XMM-Newton} \citep{Funk2007}
and \emph{Chandra} \citep{Lemiere2009}, one compact and one extended X-ray source were detected close to the center of the remnant and within the VHE emission. The compact source was suggested to be a pulsar, and HESS J1640$-$465 was first interpreted as the associated pulsar wind nebula (PWN), but no pulsation was detected from these datasets. Based on H\,{\sc I} absorption features, \cite{Lemiere2009} derived a distance of (8 -- 13) kpc, implying a large size for this potential TeV PWN. Later on, observations by the \emph{Fermi}-Large Area Telescope (LAT) revealed a high-energy $\gamma$-ray source coincident with HESS J1640$-$465 \citep[1FGL J1640.8$-$4634,][]{Slane2010} with a steep spectrum of index $\Gamma = 2.3 \pm 0.1$. These multi-wavelength data can be reproduced only in the context of a PWN origin if the electron spectrum contains a distinct low-energy component such as a Maxwellian electron population
with a power-law tail. More recently, new H.E.S.S. data showed that the VHE  $\gamma$-ray emission from HESS J1640$-$465 significantly overlaps the north-western part of the SNR shell of G338.3$-$0.0 \citep{hessj1640ohm}. In addition, the TeV spectrum connects smoothly with the
\emph{Fermi} spectrum and presents a high-energy cut-off, implying that particles with tens of TeV energies are present in the acceleration region. These new data are better explained in a scenario where protons are being accelerated at the shell of SNR G338.0$-$0.0 and $\gamma$-rays are being produced via proton-proton interaction. To render the whole story even more complex, a few months after this new H.E.S.S. publication, \cite{Gotthelf2014} reported the discovery of pulsations from the X-ray point source previously detected by \cite{Funk2007} and \cite{Lemiere2009} using the Nuclear Spectroscopic Telescope Array (NuSTAR) X-ray observatory. No pulsation was detected using 5 years of \emph{Fermi}-LAT data. However, this new detection restores interest in the scenario of a PWN detected at GeV and TeV energies. Indeed, with a spin-down luminosity of $4.4 \times 10^{36}$ erg s$^{-1}$ and a characteristic age of 3350 years, this 206 ms pulsar PSR J1640$-$4631 presents similar properties with the pulsars powering TeV PWNe~\citep{tevpwne}. 

Located only $0\fdg25$ away from HESS J1640$-$465, HESS J1641$-$463 remained unnoticed using the standard detection technique used by the H.E.S.S. Collaboration. Its detection became possible only by using an energy threshold of 4 TeV \citep{hessj1641oya}. This new TeV source presents a hard spectrum ($\Gamma$ $\sim 2$), which explains why the distinction between the two TeV sources became more evident above a few TeVs. It is coincident with the radio SNR G338.5+0.1, located at 11 kpc \citep{Kothes2007} and adjacent to SNR G338.3$-$0.0. In radio, the two remnants are connected in projection by the bright HII region G338.4+0.0. No other known counterpart is found to be compatible with HESS J1641$-$463 and no \emph{Fermi}-LAT source was reported before.

Here, we carried out a complete analysis of this region with \emph{Fermi}-LAT data accumulated over 5 years using the most recent instrument response functions (IRFs) and diffuse background models. In Section \ref{lat}, we describe the $\gamma$-ray observations used, while Section~\ref{results} presents the results obtained from a detailed morphological and spectral analysis of the \emph{Fermi}-LAT data. Finally, in Section~\ref{discussion}, we discuss the main implications of these results concerning the origin of the detected $\gamma$-ray signal. 

%%%%%%%%%%%%%%%%%%%%%%%%%%%%%%%%%%%%%%%%%%%%%%%%%%%%%%%%%%%%%%%%%%%%%%%%%%%%%%%%%%%%%%%%%%%%%%%%%%%%%%%%%%%%%
\section{LAT description and observations}
\label{lat}
The LAT is an electron-positron pair-conversion telescope, sensitive to $\gamma$-rays with energies from 20 MeV to more than 300 GeV~\citep{atwood2009}. %It is made of a high-resolution converter tracker (direction measurement of the incident $\gamma$-rays), a CsI(Tl) crystal calorimeter (energy measurement) and an anticoincidence detector to identify the background of charged particles.

The following analysis was performed using more than 5 years of \emph{Fermi}-LAT reprocessed Pass~7 data collected primarily in survey mode (2008 August 04 -- 2013 December 4). This new dataset consists of Pass 7 LAT data that have been reprocessed using updated calibration constants for the detector systems \citep{p7rep}. %The primary differences with respect to the Pass7\_V6 IRFs are the correction of a slight (1\% per year) expected degradation in the calorimeter light yield and significant improvement of the position from the calorimeter reconstruction, which in turn significantly improves the LAT point-spread function above 5 GeV. 
Only $\gamma$ rays in the Source class events were selected, excluding those coming from a zenith angle larger than 100$^{\circ}$ to the detector axis to reduce contamination from the Earth limb. We adopted the P7REP\_SOURCE\_V15 IRFs. 

%%%%%%%%%%%%%%%%%%%%%%%%%%%%%%%%%%%%%%%%%%%%%%%%%%%%%%%%%%%%%%%%%%%%%%%%%%%%%%%%%%%%%%%%%%%%%%%%%%%%%%%%%%%%%
\section{\emph{Fermi}-LAT analysis of HESS J1640$-$465 and HESS J1641$-$463}
\label{results}
The spatial and spectral analysis of the $\gamma$-ray data is performed using two different tools, $\mathtt{gtlike}$ and $\mathtt{pointlike}$.
$\mathtt{gtlike}$ is a maximum-likelihood method~\citep{Mattox1996} implemented in the Science Tools distributed by the \emph{Fermi} Science Support Center (FSSC)\footnote{http://fermi.gsfc.nasa.gov/ssc/}.
$\mathtt{pointlike}$ is an alternate binned likelihood technique, optimized for characterizing the extension of a source (unlike
$\mathtt{gtlike}$), that was extensively tested against $\mathtt{gtlike}$ \citep{Kerr2011, Lande2012}. These tools fit a model of the
region, including sources, residual charged particles, extragalactic and Galactic backgrounds, to the data. For this work, the Galactic diffuse emission model gll$\_$iem$\_$v05.fits and isotropic diffuse model iso$\_$source$\_$v05.txt provided by the \emph{Fermi}-LAT Collaboration were used\footnote{These models are available at : http://fermi.gsfc.nasa.gov/ssc/data/access/lat/BackgroundModels.html}. Sources within 20$^{\circ}$ of HESS J1640$-$465 are extracted from an internal 4-year source list and used in the likelihood fit. The spectral parameters of sources closer than 5$^{\circ}$ to HESS J1640$-$465 (located at the center of our square region of interest of 10$\degr$ half-side) are left free, while the parameters of all other sources are fixed at the values from the 4-year list.

\subsection{Morphological analysis}
\label{morpho}
\begin{figure}
\begin{center}
\includegraphics[width=0.45\textwidth]{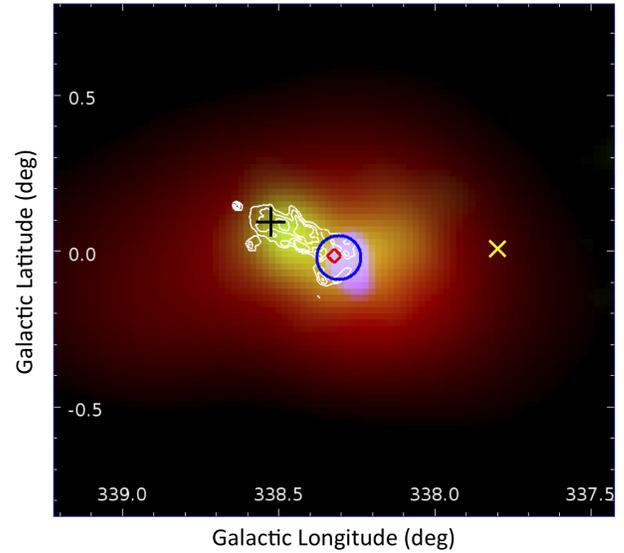}
\caption{\emph{Fermi}-LAT composite TS map of the region containing the two TeV sources in the 0.3 -- 3.0 (red), 3.0 -- 30.0 GeV (green) and 30.0 -- 300.0 GeV (blue) energy ranges, in Galactic coordinates. The best position of the H.E.S.S. point source HESS~J1641$-$463 is marked with a black plus while the best position and extension of the H.E.S.S. Gaussian HESS~J1640$-$465 is represented by a blue circle. The red diamond indicates the position of the pulsar PSR J1640$-$4631, while the yellow cross shows the position of the only close-by source from the internal \emph{Fermi}-LAT source list located in the region. Radio continuum at 843~MHz is shown as white contours \citep{whiteoak96}. The maximal TS values are 255, 100, 43 for the red, green and blue maps respectively.}
\label{fig:TSmap}
\end{center}
\end{figure}

Two sources are detected in the 4-year list, positionally coincident with HESS J1640$-$465 and HESS J1641$-$463. To have a better understanding of this complex region, we first generated residual Test Statistic (TS) maps in three different energy bands using $\mathtt{pointlike}$: 0.3 -- 3 GeV, 3 -- 30 GeV, 30 -- 300 GeV. The low energy threshold of 0.3~GeV was defined as a good compromise between statistics and angular resolution.  $TS = 2 (\log \mathcal{L}_1 - \log \mathcal{L}_0)$ where $\mathcal{L}_1$ is the likelihood obtained by fitting the source plus background model to the data, and $\mathcal{L}_0$ is obtained by fitting the background alone~\citep{Mattox1996}. For each energy band, this skymap contains the TS value for a point source at  each map location assuming a fixed photon index $\Gamma$ = 2, thus giving a measure of the statistical significance for the detection of a $\gamma$-ray source in excess of the background. The diffuse Galactic and isotropic emission, as well as nearby sources (except the two sources coincident with our two TeV sources of interest) are included in the background model. The resulting RGB composite TS map (Figure~\ref{fig:TSmap}) clearly shows a $\gamma$-ray signal coincident with the position of HESS J1641$-$463 in the 3 -- 30 GeV energy range, while the only remaining $\gamma$-ray emission in the 30 -- 300 GeV energy band is coincident with HESS J1640$-$465. This is a very good indication that, indeed, two components are visible with \emph{Fermi}-LAT and that the $\gamma$-ray emission coming from HESS J1640$-$465 may be harder. 

\begin{deluxetable*}{rlccccc}
\tablecaption{Centroid and extension fits of the LAT data using $\mathtt{pointlike}$ above 3 GeV. \label{table:centroid_pointlike} }
\tablewidth{0pt}
\tablehead{ 
~ & \colhead{Spatial Model} & \colhead{R.A. ($^\circ$)} & \colhead{Dec. ($^\circ$) } & $\sigma$ ($^\circ$) & \colhead{TS (3 -- 300 GeV)} & \colhead{N.d.o.f.**} 
}
\startdata
$A$ & 1 Point Source 	& 250.14 & $-$46.56 & & 148 & 4 \\
$B$ & 1 Gaussian Source & 250.17 & $-$46.52 & 0.12 & 186 & 5 \\
$C$ & 1 Gaussian Source with 2 power-law type components & 250.17  & $-$46.52  & 0.12 & 191 & 7 \\
$D$ & 2 Point Sources & 250.13 / 250.27 & $-$46.57 / $-$46.35 & & 216 & 8 \\
\vspace{0.2cm} 
$E$ & 1 Gaussian + 1 Point Source & 250.15 / 250.27 & $-$46.59 / $-$46.34 & 0.06 / -- & 222 & 9 \\		
$F$ & HESS J1640$-$465 Gaussian* & 250.17 & $-$46.54 & 0.07 & 166 & 2 \\
$G$ & HESS J1640$-$465 Gaussian* + HESS J1641$-$463 Point Source* & 250.17 / 250.26 & $-$46.54 / $-$46.30 & 0.07 / -- & 208 & 4 \\
\enddata
\tablenotetext{*}{H.E.S.S. positions and extensions are fixed to the values from \cite{hessj1640ohm} and \cite{hessj1641oya}.} 
\tablenotetext{**}{N. d.o.f. : number of degrees of freedom.} 
\end{deluxetable*}

In a second step, to ensure that the two sources are significantly detected and check their possible extension, we used $\mathtt{pointlike}$ above 3 GeV, thus only keeping events with the best angular reconstruction. We tested different hypotheses to represent the emission coming from HESS J1640$-$465 and HESS J1641$-$463: one point source (model $A$), one Gaussian distribution (model $B$), one Gaussian distribution with two power-law type components (model $C$), two point sources (model $D$), and one Gaussian plus one point source (model $E$). We also used the best spatial shape (1-D Gaussian) representing HESS J1640$-$465 as observed by H.E.S.S. \citep[model $F$,][]{hessj1640ohm}, and another spatial model composed of this best Gaussian distribution and the best-fit position of a point source for HESS J1641$-$463 \citep[model $G$, also obtained by H.E.S.S., ][]{hessj1641oya}. For all these tests, power-law spectral shapes were assumed. The results of the extension fits and the improvement of the TS when using spatially extended models are summarized in Table~\ref{table:centroid_pointlike}, along with the number of additional degrees of freedom with respect to the null hypothesis. The $5\sigma$ improvement of the likelihood fit between a Gaussian distribution and the Gaussian plus point-source hypothesis (difference in TS of 36 for 4 additional degrees of freedom) supports the conclusion that two different components are detected in this region and, above all, that a significant $\gamma$-ray source is located in the vicinity of HESS J1641$-$463. This is the first time that a detection of this source is reported at these energies. The improvement when using a Gaussian plus a point source instead of the two point-source hypothesis is marginal ($2.4\sigma$) and therefore one cannot exclude the possibility that both sources are point-like in the GeV energy range. However, in both hypotheses ($D$ and $E$), the best-fit positions obtained with the \emph{Fermi}-LAT data are in good agreement with those extracted from H.E.S.S. analyses \citep{hessj1640ohm, hessj1641oya}. In model $E$, the best-fit position for the Gaussian is $\alpha(\rm{J2000})=250\fdg15$ , $\delta(\rm{J2000})=-46\fdg59$ with a 68\% error radius of $0\fdg02$, and an extension of $0\fdg06$ $\pm 0\fdg02$. The best-fit position for the additional point source is $\alpha(\rm{J2000})=250\fdg27$ , $\delta(\rm{J2000})=-46\fdg34$ with a 68\% error radius of $0\fdg02$. Since this model ($E$) is not significantly better than the same model with parameters fixed to the values extracted from \cite{hessj1640ohm} and \cite{hessj1641oya}, at the expense of a larger number of degrees of freedom, we decided to use the H.E.S.S. spatial shapes  (model $G$) to derive the spectra of both sources.

%
%\subsection{Timing analysis}
%\label{timing}
%The morphological analysis allowed the detection of two distinct and significant gamma-ray sources. To ensure that these two sources are not contaminated by pulsed photons from the recently discovered pulsar PSR J1640$-$4631, we carried a timing analysis. Indeed, with its large spin-down power of $4.4 \times 10^{36}$ erg s$^{-1}$,  PSR J1640$-$4631 is similar to other middle-aged pulsars powering TeV nebulae.
%This pulsar is not monitored as part of the LAT pulsar timing campaign \citep{Smith 2008}, as it was discovered subsequently. Therefore, we used the timing solution reported in \cite{Gotthelf2014}. \emph{Fermi}-LAT photons with energies above 100 MeV and within a radius of 1.0$^{\circ}$ of the radio pulsar position $\alpha(J2000) = 16^{h}40^{m}43.52^{s}$, $\delta(J2000) = -46\degr31'35.4"$ were selected using an energy-dependent cone of radius $\displaystyle \theta_{68}<\max(5.12^{\circ} \times (E/100 \, {\rm MeV})^{-0.8},0.2^{\circ})$  and phase-folded using this radio ephemeris. This choice takes into account the instrument performance and improves the signal to noise ratio over a broad energy range. No significant pulsation was detected for all tested energy band (100 MeV -- 300 GeV, 100 MeV -- 300 MeV, 300 MeV -- 1 GeV, $>$ 1 GeV). This lack of significant gamma-ray pulsations is in agreement with the work reported in \cite{Gotthelf2014}.
%

\subsection{Spectral analysis}
\label{spec}
The following spectral analysis was performed with $\mathtt{gtlike}$ using all events between 0.1 and 300 GeV. As discussed in Section~\ref{morpho}, we used the spatial model $G$ from Table~\ref{table:centroid_pointlike} to represent the $\gamma$-ray emission observed by the LAT from HESS J1640$-$465 and HESS J1641$-$463. Assuming this spatial shape, the $\gamma$-ray sources observed by the LAT above 100 MeV are detected with a TS of 148 for HESS J1640$-$465 and 105 for HESS J1641$-$463. Their spectra are presented in Figures~\ref{fig:Spectra1} and \ref{fig:Spectra2}. They are both well described by simple power laws (dN/dE $\propto$ E$^{-\Gamma}$) with spectral indices of $\Gamma = 1.99 \pm 0.04 \pm 0.07$ for the source HESS J1640$-$465 and $\Gamma = 2.47 \pm 0.05 \pm 0.06$ for the source HESS J1641$-$463. This confirms the trend of a harder spectrum for HESS J1640$-$465, as suggested by Figure~\ref{fig:TSmap} and further discussed in Section~\ref{morpho}. The integrated flux above 100 MeV is $(4.5 \pm 0.2 \pm 0.2) \times 10^{-5}$ MeV cm$^{-2}$ s$^{-1}$ for HESS J1640$-$465 and $(2.6 \pm 0.2 \pm 0.5) \times 10^{-5}$ MeV cm$^{-2}$ s$^{-1}$ for HESS J1641$-$463. The first error is statistical, while the second represents our estimate of systematic effects as discussed below. No evidence of variability was found for these two sources by dividing the data into yearly time bins and applying the likelihood procedure for each.

The \emph{Fermi}-LAT spectral points shown in Figures~\ref{fig:Spectra1} and \ref{fig:Spectra2} were obtained by dividing the 0.1~--~300 GeV range into 14 logarithmically spaced energy bins and performing a maximum likelihood spectral analysis in each interval, assuming a power-law shape with fixed photon index $\Gamma$ = 2 for both sources. The normalizations of the diffuse Galactic and isotropic emission were left free in each energy bin as well as those of sources located within $5^{\circ}$ from HESS J1640$-$465. A 95\% C.L. upper limit is computed when the statistical significance is lower than 2$\sigma$. At the highest energies, bins corresponding to upper limits were combined. The errors on the spectral points represent the statistical and systematic uncertainties added in quadrature. 

Two main systematic errors have been taken into account: imperfect modeling of the Galactic diffuse emission and uncertainties in the effective area calibration. The first was estimated by comparing the results obtained using the standard Galactic diffuse model with the results based on eight alternative interstellar emission models as performed in \cite{snrcat}. The second is estimated by using modified IRFs whose effective areas bracket the nominal ones. These bracketing IRFs are defined by envelopes above and below the nominal energy dependence of the effective area by linearly connecting differences of (10\%, 5\%, 5\%, 15\%) at $\log_{10}(E/{\rm MeV})$ of (2, 2.5, 4, 6), respectively\footnote{http://fermi.gsfc.nasa.gov/ssc/data/analysis/LAT\_caveats.html}.

%%%%%%%%%%%%%%%%%%%%%%%%%%%%%%%%%%%%%%%%%%%%%%%%%%%%%%%%%%%%%%%%%%%%%%%%%%%%%%%%%%%%%%%%%%%%%%%%%%%%%%%%%%%%%%%%%%%%%%%%%%%%%%%%%%%
\section{Discussion}
\label{discussion}
Thanks to a larger dataset and to an improved event reconstruction, \emph{Fermi}-LAT analysis of the region containing the two TeV sources HESS J1640$-$465 and HESS J1641$-$463 has led to the detection of two distinct GeV sources of $\gamma$-ray photons. The first one, coincident with HESS J1640$-$465 presents a hard spectrum, while the second one shows a much steeper spectrum. These new characteristics can be used to constrain the natures of these two $\gamma$-ray sources. In the following, we will assume that HESS J1640$-$465 is located at 10~kpc \citep[as in][]{hessj1640ohm} and that HESS J1641$-$463 is 11 kpc from us \citep[following][]{hessj1641oya}.

\begin{figure}[h!!]
\begin{center}
\includegraphics[width=0.45\textwidth]{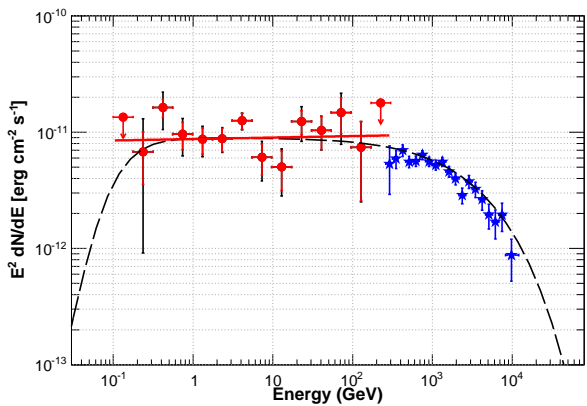}
\caption{Gamma-ray spectra of HESS J1640$-$465, using the best Gaussian spatial model from \cite{hessj1640ohm}. The red line shows the fit of a power-law to the \emph{Fermi} spectrum derived above 100 MeV. The red data points (crosses) indicate the fluxes measured in each of the 14 energy bins indicated by the extent of their horizontal lines. The statistical errors are shown in red, while the black lines take into account both the statistical and systematic errors as discussed in Section~\ref{spec}.  A 95\% C.L. upper limit is computed when the statistical significance is lower than 2$\sigma$. The blue data points are taken from \cite{hessj1640ohm}. The dashed black line indicates the hadronic model discussed in Section~\ref{discussion}.}
\label{fig:Spectra1}
\end{center}
\end{figure}

\begin{figure}[h!!]
\begin{center}
\includegraphics[width=0.45\textwidth]{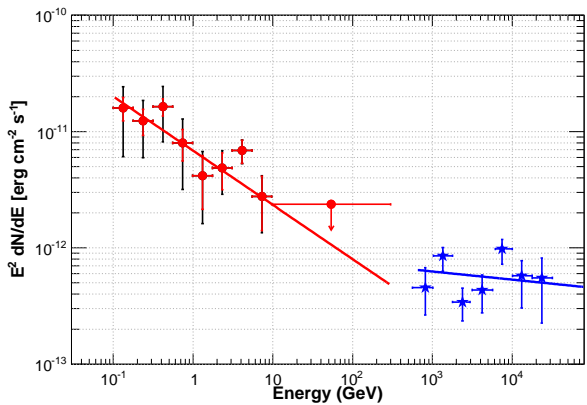}
\caption{Gamma-ray spectra of HESS J1641$-$463, using the best point source position from \cite{hessj1641oya}. Same conventions as for Figure~\ref{fig:Spectra1}. The blue data points and the best spectral fit (represented as a blue line) are taken from \cite{hessj1641oya}.}
\label{fig:Spectra2}
\end{center}
\end{figure}

For the case of HESS J1640$-$465, the hard spectrum revealed in this new analysis connects perfectly with the TeV spectral points (see Figure~\ref{fig:Spectra1}). The resulting $\gamma$-ray luminosity above 100 MeV for this source is $L_{\ge 100 MeV} \approx (6.8 \pm 0.6) \times 10^{35}$ ($D / 10$ kpc)$^2$ erg s$^{-1}$; this is three times less than reported by \cite{Slane2010}. This significant change is mainly due to the large contamination below 1 GeV implied by the close-by source HESS J1641$-$463, which shows a much steeper spectrum in this energy band. The high $\gamma$-ray luminosity and flat spectral shape of HESS J1640$-$465 is typical of SNRs interacting with molecular clouds previously detected with \emph{Fermi}, such as IC~443 and W51C, which both have luminosities $\ge$10$^{35}$ erg s$^{-1}$ \citep{fermi:ic443,fermi:w51c}. The new \emph{Fermi}-LAT data thus strengthen the SNR scenario discussed in \cite{hessj1640ohm} where protons are accelerated in the shell of SNR G338.3$-$0.0 and the $\gamma$-ray emission from HESS J1640$-$465 arises from the interaction region between the blast wave and the G338.4+0.1 H~{\sc II} complex. In the framework of such hadronic scenario, we calculated the flux and spectrum of the $\gamma$ rays produced by $\pi^0$-decay following the analytical approximations by \cite{Kelner2006}, assuming a gas density of 150 cm$^{-3}$ \citep{hessj1640ohm}. The particle index is fixed at 2.0 (in agreement with the spectral index derived in this work) with an energy cut-off at 50 TeV taking into account the spectral curvature observed by H.E.S.S. at TeV energies. The total energy in cosmic-ray (CR) hadrons required to reproduce the \emph{Fermi}-LAT and H.E.S.S. data (see dashed line in Figure~\ref{fig:Spectra1}) is $\frac{0.8}{f}\times 10^{50}$ erg, where $f$ corresponds to the fraction of the SNR that is interacting with the molecular cloud. In case of full immersion ($f = 1$) and for a typical explosion energy of $\sim10^{51}$ erg, this hadronic model indicates that only $\sim 8$\% of the initial kinetic energy needs to be transferred to CRs. Since $f$ is a poorly known fraction, the derived energy should be considered as a lower limit. The high density needed and the high-energy cut-off visible in this source contrast with other \emph{Fermi}-LAT detected SNRs. 
%thus ranging HESS J1640$-$465 as a standard SNR interacting with molecular clouds. 

One cannot exclude the possibility that part of the flux detected by \emph{Fermi} is produced by PSR J1640$-$4631 or even by its PWN in SNR G338.3$-$0.0, which would further decrease these values. In the case of contamination by PSR J1640$-$4631, though no pulsation was detected by \cite{Gotthelf2014} in the \emph{Fermi}-LAT data, the absence of spectral curvature or cut-off up to 300 GeV in the \emph{Fermi} source implies that the bulk of the emission does not arise from this pulsar, especially above 10 GeV. The PWN scenario is more likely, and it is interesting to note that the PWN modeling presented in \cite{Gotthelf2014} reproduces relatively well the high-energy end of the spectrum obtained in our new analysis (above a few GeV). In this context, the marginally significant extension indicated by our analysis would be an important argument in favor of the PWN or the SNR scenario, if it is confirmed with more data.

In the case of HESS J1641$-$463, the connection between the steep spectrum detected by this new analysis and the H.E.S.S. spectrum remains unclear and points toward two different mechanisms or sources producing the $\gamma$-ray photons detected in each energy band. This recalls the case of HESS J1356$-$645, within which \emph{Fermi}-LAT detects the pulsed emission of PSR~J1357$-$6429 while, at TeV energies, H.E.S.S. observes its associated PWN \citep{hessj1356}. In such a PSR/PWN scenario, the flux density of the X-ray source nearest to HESS~J1641$-$463 is a factor 15 less energetic when compared with the energy flux density of the TeV source \citep{hessj1641oya}, which would resemble the numerous associations of ÒdarkÓ TeV sources and weak X-ray synchrotron PWN that have been established so far. Another possibility in such a "two source" scenario, emphasized by the good positional coincidence between HESS J1641$-$643 with SNR G338.5+0.1, would be that the $\gamma$-ray photons detected at TeV energies are produced by accelerated protons from this SNR interacting with the ambient gas, as suggested by \cite{hessj1641oya}, while \emph{Fermi} would be seeing a contaminating pulsar. Finally, another scenario would be that HESS J1641$-$463 is a binary system similar to LS~5039 and LSI +61$\degr$303, both showing a pulsar-like spectrum at GeV energies that does not connect to the TeV spectrum \citep{Hadasch2012}. In this regard, the non detection of spectral curvature and/or variability with the current \emph{Fermi}-LAT dataset does not permit excluding this last scenario since the GeV source is rather weak and the statistics are too low to perform a blind search for pulsation or to divide the dataset into intervals shorter than one year. 
  
Further observations as well as timing and flux variability searches with \emph{Fermi} and H.E.S.S. and also with high-resolution X-ray instruments will be key to finally understanding the nature of these two intriguing $\gamma$-ray sources.

%%%%%%%%%%%%%%%%%%%%%%%%%%%%%%%%%%%%%%%%%%%%%%%%%%%%%%%%%%%%%%%%%%%%%%%%%%%%%%%%%%%%%%%%%%%%%%%%%%%%%%%%%%%%%%%%%%%%%%%%%%%%%%%%%%%

\acknowledgments
The \textit{Fermi} LAT Collaboration acknowledges generous ongoing support
from a number of agencies and institutes that have supported both the
development and the operation of the LAT as well as scientific data analysis.
These include the National Aeronautics and Space Administration and the
Department of Energy in the United States, the Commissariat \`a l'Energie Atomique
and the Centre National de la Recherche Scientifique / Institut National de Physique
Nucl\'eaire et de Physique des Particules in France, the Agenzia Spaziale Italiana
and the Istituto Nazionale di Fisica Nucleare in Italy, the Ministry of Education,
Culture, Sports, Science and Technology (MEXT), High Energy Accelerator Research
Organization (KEK) and Japan Aerospace Exploration Agency (JAXA) in Japan, and
the K.~A.~Wallenberg Foundation, the Swedish Research Council and the
Swedish National Space Board in Sweden.
 
Additional support for science analysis during the operations phase from the following agencies is also gratefully acknowledged: the Istituto Nazionale di Astrofisica in Italy and the Centre National d'\'Etudes Spatiales in France.

%% The following command ends your manuscript. LaTeX will ignore any text
%% End of file 

\end{document}